\documentclass[runningheads]{llncs}
\usepackage[T1]{fontenc}
\usepackage{booktabs}
\usepackage{graphicx}
\usepackage{graphicx}
\usepackage{bibunits}
\usepackage{multirow}
\usepackage[misc]{ifsym}
\usepackage{booktabs}
\usepackage{amsmath}
\usepackage{amsmath,amssymb,amsfonts}
\usepackage{bm}
\usepackage{tabularray}
\usepackage{url}

\usepackage{authblk}
\usepackage[table]{xcolor}

\newcommand{\ie}{\textit{i.e.}}
\newcommand{\eg}{\textit{e.g.}}

\definecolor{LightCyan}{RGB}{200,220,240}
\definecolor{myPink}{RGB}{240,212,220}

\title{LatentArtiFusion: An Effective and Efficient Histological Artifacts Restoration Framework}

\titlerunning{LatentArtiFusion}
\author{
Zhenqi He$^{\star}$ \and Wenrui Liu\thanks{Equal contribution. The order is random.} \and Minghao Yin \and Kai Han\textsuperscript{(\Letter)}
}

\authorrunning{Z. He \& W. Liu  et al.}
\institute{The University of Hong Kong\\
\email{kaihanx@hku.hk}
}

\begin{document}

\maketitle

\begin{abstract}

Histological artifacts pose challenges for both pathologists and Computer-Aided Diagnosis (CAD) systems, leading to errors in analysis. Current approaches for histological artifact restoration, based on Generative Adversarial Networks (GANs) and pixel-level Diffusion Models, suffer from performance limitations and computational inefficiencies. In this paper, we propose a novel framework, LatentArtiFusion, which leverages the latent diffusion model (LDM) to reconstruct histological artifacts with high performance and computational efficiency.
Unlike traditional pixel-level diffusion frameworks, LatentArtiFusion executes the restoration process in a lower-dimensional latent space, significantly improving computational efficiency. 
Moreover, we introduce a novel regional artifact reconstruction algorithm in latent space to prevent mistransfer in non-artifact regions, distinguishing our approach from GAN-based methods.
Through extensive experiments on real-world histology datasets, LatentArtiFusion demonstrates remarkable speed, outperforming state-of-the-art pixel-level diffusion frameworks by more than $30\times$. It also consistently surpasses GAN-based methods by at least 5\% across multiple evaluation metrics. Furthermore, we evaluate the effectiveness of our proposed framework in downstream tissue classification tasks, showcasing its practical utility. Code is available at \url{https://github.com/bugs-creator/LatentArtiFusion}.
\end{abstract}


\section{Introduction}
\parskip=0pt
Histological Whole Slide Images (WSI) provide ample information on tissue and nuclei structures, serving as a valuable resource for contemporary clinical diagnosis and informing treatment decisions~\cite{histology,transnuseg}. Both pathologists and Computer-Aided Diagnosis (CAD) rely on the analysis of morphological and contextual information present in histological images. Nevertheless, the intricate digitization process and potential mishandling can commonly lead to alterations in tissue structure or staining information, such as tissue folding, bubbles, tissue mask, and other deformation~\cite{artifacts,reviewArtifacts,artifacttype}. 
Such deformation of tissue structures, namely artifacts in the context of pathology, not only poses challenges for both pathologists and Computer-Aided Diagnosis but also elevates 
the risk of misdiagnosis, \ie, incorrectly identifying artifacts as tumors~\cite{artifactresult}.
Many existing methods struggle to achieve satisfactory results in the presence of histological artifacts~\cite{zhang2022benchmarking}.

To address the challenge posed by the histological artifacts, many methods have been proposed based on Generative Adversarial Networks (GANs). For example,
AR-CycleGAN~\cite{arcyclegan} employs a Cycle Generative Adversarial Network (CycleGAN)~\cite{CycleGAN2017,gan} to learn the transfer between unpaired artifact and artifact-free images through adversarial learning. This model frames artifact restoration as a domain transfer problem, aiming to transform artifact images into clean images. However, this approach introduces a potential challenge of unintended stain style transfer in non-artifact regions, thereby increasing the risk of misdiagnosis. 
Additionally, GAN-based models are prone to failures during training due to mode collapse and mode dropping~\cite{GANoverview}, posing challenges for their practical use. 
Recently, with the success of diffusion models~\cite{DDPM,song2021scorebased}, ArtiFusion~\cite{artifusion} is introduced, which leverages the denoising diffusion probabilistic model~\cite{DDPM} to selectively reconstruct the artifact region, aiming to preserve the structural and stain information of non-artifact areas. Further, a lightweight transformer based denoising backbone is employed for efficient histological artifact restoration~\cite{wang2024lightweight}. While, the restorations is still taken by the denoising process of DDPM within the pixel-level space, which, given its substantial size, results in considerable computational costs.
%
%
For example, DDPM based methods take approximately 30 times longer processing time compared to GAN-based methods, making it inefficient for real applications.

In this paper, we aim to achieve high performance on histological artifacts restoration while maintaining high computational efficiency. To this end, we propose \textbf{LatentArtiFusion}, in which we seamlessly integrate Variational AutoEncoder (VAE)~\cite{VAE} and the Denoising Diffusion Probabilistic Model (DDPM)~\cite{DDPM}.
This combination enables a denoising process within a perceptually equivalent space with lower-dimensions, thereby preserving the high performance of DDPM and achieving comparably lower computational costs. Moreover, a latent regional denoising process is proposed to simultaneously leverage and preserve the semantic information of adjacent non-artifact regions, specifically tailored for artifact restoration. 

Our major contributions are: (1) We present a framework for histological artifact restoration, as the first attempt at latent diffusion models driven framework for histological artifact restoration. This approach opens up new possibilities for artifact restoration in histopathology. (2) We propose a novel regional latent denoising algorithm guided by artifact masks, allowing selective reconstruction of the artifact regions. This algorithm effectively preserves the stain style of non-artifact regions, resulting in visually consistent and coherent restorations. (3) Our method demonstrates superior efficiency, significantly surpassing previous pixel-level diffusion frameworks (\eg, by over 30 times faster compared with ArtiFusion) in processing speed. Additionally, we achieve state-of-the-art results across multiple datasets, highlighting the superior performance and effectiveness of our approach.

\section{Method}

\begin{figure}[t]

  \includegraphics[width=\linewidth]{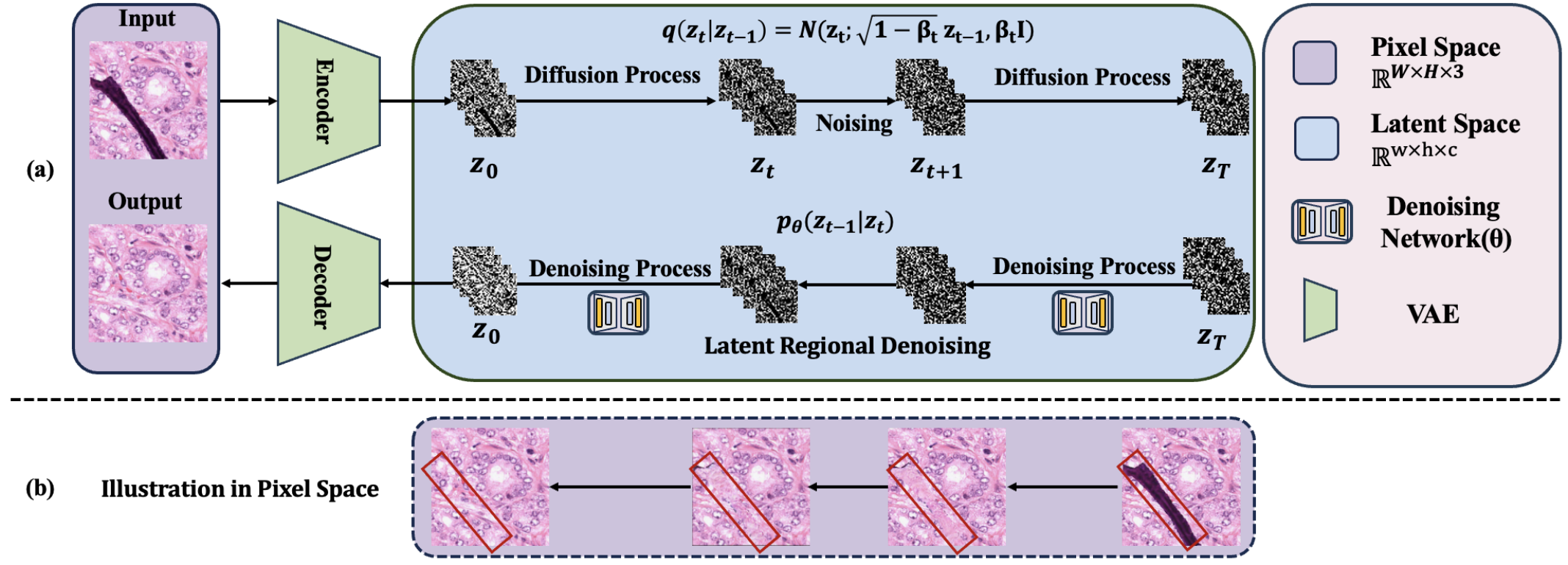}
  \caption{
  (a) Overall architecture of the proposed \textbf{LatentArtiFusion} framework. (b) Visualization of latent regional denoising process in pixel-level space (Artifact region is marked with \textcolor{red}{$\square$}).
  }
  \label{fig:model}
\end{figure}
\subsection{Preliminary in Diffusion Models}
\label{sec:DM}

\subsubsection{Denoising Diffusion Probabilistic Model (DDPM)~\cite{DDPM}. } DDPM learns the complex data distribution through diffusion and denoising process. In diffusion process, given the input data $X_0 \sim p(X)$, random Gaussian noise is gradually added into $X_0$ to obtain the noisy version $X_t$ for timestep $t \in \mathbb{N}[0,T]$ following the Markov process $q(\mathbf{x}_t|\mathbf{x}_{t-1}) = \mathcal{N}(\mathbf{x}_{t};\sqrt{1-\beta_t}\mathbf{x}_{t-1},\beta_t\mathbf{I})$, where $T$ is the total diffusion timesteps , $X_T \sim N(0,\bm{I}) $ represents white noise without any semantic information, and $\beta_t$ is the hyper-parameter representing the variance schedule. In the denoising process, a U-shaped denoising network with parameter $\theta$ is trained to reverse the diffusion process $q(\mathbf{x}_t|\mathbf{x}_{t-1})$ to learn $p_{\theta}(\mathbf{x}_{t-1}|\mathbf{x}_{t}^{in})$ and gently denoise $X_T$ back to $X_0$ to recover the information of data $X_0$. The overall training objectives are formulated by the variational lower bound of the negative log-likelihood and can be written as:

\begin{equation}
\small
\nonumber
    L=
    \mathbb{E}_{q}
    [\underbrace{D_{KL}(q(\mathbf{x}_{T}|x_{0}))||p(\mathbf{x}_{T})}_{L_{T}} 
    + \sum_{t>1}
    \underbrace{D_{KL}(q(\mathbf{x}_{t-1}|\mathbf{x}_{t},\mathbf{x}_{0}))||p_{\theta}(\mathbf{x}_{t-1}|\mathbf{x}_{t})}_{L_{t-1}}-\underbrace{\log p_{\theta}(\mathbf{x}_{0}|\mathbf{x}_{1})}_{L_{0}}],
\end{equation}
where $D_{KL}(\cdot||\cdot)$ is the KL divergence.

\subsubsection{Latent Diffusion Models (LDM)~\cite{latentdiffusion}.}
Denoising in large data space, \ie, pixel-level space $\mathbb{R}^{W \times H \times 3}$, is relatively computationally costly and inefficient. To reduce the computational costs, 
LDM conducts the diffusion and denoising process in a lower-dimensional latent space through compressing data $X$ through Variational Auto-Encoder $\varepsilon (\cdot )$ (VAE)~\cite{VAE} structure into a smaller dimensional space $\mathbb{R}^{w \times h \times c}$, and output the decoded $\hat{X} = D(\varepsilon (X ))$, where $\varepsilon$ and $D$ are the encoder and decoder of the VAE.

\begin{figure}[h!]
\centering
\includegraphics[width=0.6\textwidth]{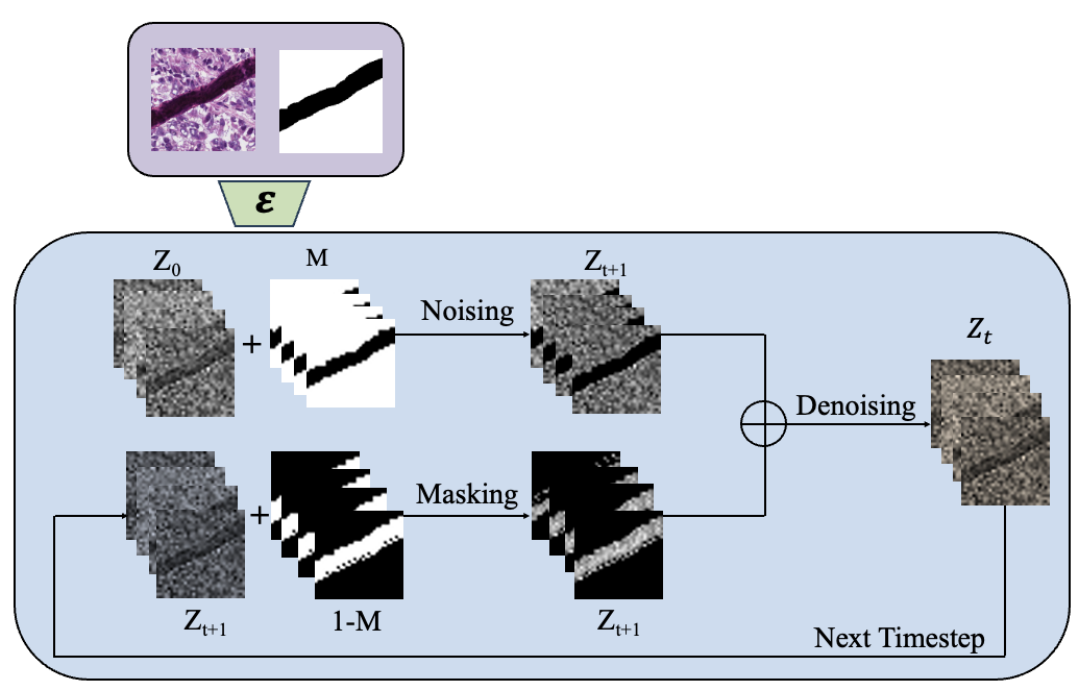}
\caption{Regional denoising in latent space to selectively denoising the artifact regions while preserving the clean regions.
}
\label{fig:inference}
\end{figure}

\subsection{LatentArtiFusion}
\subsubsection{Overall Pipeline. }The comprehensive architecture of our proposed framework is illustrated in Fig.~\ref{fig:model}(a). The framework comprises two key stages: training and inference, constituting the backbone of our restoration approach.
During the training stage, our primary objectives involve the training of a latent diffusion model. This model assimilates contextual information and captures the color distribution inherent in real artifact-free histology images.
In the inference stage, We propose a novel latent regional denoising algorithm to selectively reconstruct artifact regions in latent space leveraging the semantic information of nearby artifact-free regions to preserve the morphological details of non-artifact regions and maintain visual consistence in reconstructed artifact regions. Fig.~\ref{fig:model}(b) visually represents the latent regional denoising process in pixel-level space to demonstrate the gradual reconstruction of the artifact region.

\subsubsection{Model Training.} For efficiency, we follow the formulation of the latent diffusion model to train our model from scratch in a lower-dimensional latent space. Initially, we utilize a pre-trained Variational Autoencoder (VAE)~\cite{VAE} to compress the input image $X \in \mathbb{R}^{W \times H \times 3}$ into the latent space $\mathbb{R}^{w \times h \times c}$. Subsequently, we perform the diffusion and denoising processes within this latent space, aiming to enhance computational efficiency. Given a compressed image $Z_0 = \varepsilon(X) \in \mathbb{R}^{w \times h \times c}$, random Gaussian noises are gruadually injected following the Markov Chain $q(\mathbf{z}_t|\mathbf{z}_{t-1}) = \mathcal{N}(\mathbf{z}_{t};\sqrt{1-\beta_t}\mathbf{z}_{t-1},\beta_t\mathbf{I})$, as shown in Fig.~\ref{fig:model} (a). Subsequently, we train a U-shaped denoising network,  which takes $\mathbf{z}_{t}$ conditioned on the time $t$ as input, to predict the corresponding noise added from $\mathbf{z}_{t-1}$ to $\mathbf{z}_{t}$ in the denoising process. This allows us to learn $p_{\theta}(\mathbf{z}_{t-1}|\mathbf{z}_{t})$ and sample the noisy data back to input data $\mathbf{z}_{0}$ step by step.

Crucially, we utilize the pre-trained VAE~\footnote{Pretrained weights are downloaded from \url{https://huggingface.co/runwayml/stable-diffusion-v1-5}.} by~\cite{latentdiffusion}, and all parameters of the VAE are frozen throughout the model training process.

\subsubsection{Regional Denoising in Latent Space.} 
During the inference stage, we introduce an innovative latent regional denoising algorithm. This algorithm selectively focuses on reconstructing the artifact region in the latent space, deviating from the conventional approach of reconstructing the entire image. As shown in Fig.~\ref{fig:inference},
the compressed histology image with artifacts and its corresponding artifact mask in the latent space retain the semantic information of artifacts, we leverage the compressed mask to precisely guide the restoration process, focusing exclusively on the artifact regions. 
To perform regional denoising in the artifact region while preserving non-artifact regions in the latent space, we encode the Boolean mask representing the localization of the artifact region into the latent space (denoted as $M$). At each denoising timestep $t$, the input to the denoising network ($\theta$) is a concatenation of the noised non-artifact regions (occurring $t$ times) with the denoised artifact region obtained from the denoising network at the previous step ($t+1$). The whole latent regional denoising process can be formulated as follows:
\begin{equation}
\begin{split}
    \centering
    & \mathbf{Z}_{t} = \texttt{Denoising} [
    \mathbf{Z}_{t+1}^{sample} \odot \mathbf{M}
    + 
   \mathbf{Z}_{t+1}^{output} \odot (1 - \mathbf{M})],\\
   & \mathbf{Z}_{t+1}^{sample} = \sqrt{\bar{a_t}} \mathbf{Z}_{0} + \sqrt{1- \bar{a_t}}\epsilon_t,
    \label{eq:inference}
\end{split}
\end{equation}
where $\mathbf{Z}_{t}^{sample}\odot  \mathbf{M}$ is non-artifact region diffused as $t$ times from the compressed input data $Z_0$, following the forward Gaussian process with $\bar{\alpha}_t=\prod_{i=1}^t(1-\beta_i)$. Additionally, $\mathbf{Z}_{t+1}^{output}$ denotes the output of the denoising network at previous timestep \ie, $p_{\theta}(\mathbf{z}_{t+1}^{output}|\mathbf{z}_{t+2})$, and the $\mathbf{Z}_{t}$ is used as $\mathbf{Z}_{t}^{output}$ in subsequent time step. Through iteration, the reconstructed sample can be obtained as $\mathbf{\hat{Z}_{0}} = \texttt{Denoising} [\mathbf{Z}_{1}^{sample} \odot \mathbf{M} + \mathbf{Z}_{1}^{output} \odot (1 - \mathbf{M})]$.

\section{Experiments}
\label{sec:exp}

\begin{table*}[t!]
\centering
\caption{
Quantitative comparison of the proposed method with CycleGAN and ArtiFusion~\cite{artifusion} on artifact restoration performance in terms of the similarities between original images and restored images. 
The best performance for each indicator is highlighted in \textbf{boldface}. $\downarrow$ denotes lower is better; $\uparrow$ denotes higher is better. 
}
\label{tab:result}
\begin{tabular}{l|l|c|c|c|c|c}
\toprule
Dataset & Methods & L2~($\times10^4$) $\downarrow$ & MSE~$\downarrow$ & SSIM~$\uparrow$ & PSNR~$\uparrow$ & FSIM~$\uparrow$ \\ 
\hline
\multirow{3}{*}{Histology} 
&CycleGAN~\cite{CycleGAN2017} & $1.8930$ & $0.5936$ &  $0.9622$     &  $42.12$  & $0.7162$     \\
& ArtiFusion~\cite{artifusion}    &   $0.4940$   & $0.2465$  &  $\bm{0.9860}$   & $\bm{48.08}$ & $0.8216$ \\
& \textbf{Ours}  & $\bm{0.4493}$ & $\bm{0.2320}$ &  $0.9810$     &    $46.92$   & $\bm{0.8913}$ \\
\hline
\multirow{3}{*}{MoNuSAC2020} &CycleGAN~\cite{CycleGAN2017} & $13.81$ & $7.250$ &  $0.8876$     &  $40.04$  & $0.7094$     \\
& ArtiFusion~\cite{artifusion}    &   $4.034$   & $2.117$  &  $0.9325$   & $40.16$ & $0.7412$ \\
& \textbf{Ours}  & \bm{$3.930$} & \bm{$2.063$} &  $\bm{0.9505}$     &    $\bm{42.01}$   & $\bm{0.8752}$    \\

\bottomrule
\end{tabular}
\end{table*}

\begin{table}[h!]
\centering
\caption{
Comparison of model efficiency in terms of (1) the averaged inference time for each image with the size of $256 \times 256 \times 3$ and (2) the number of required steps for the diffusion process (Not applicable for CycleGAN).
}
\label{tab:cost}
\begin{tabular}{l|c|c}
\toprule
Methods   & Inference(s) & Diffusion steps \\ 
\hline
CycleGAN~\cite{CycleGAN2017} &   $1.065$ &  $N/A$ \\
ArtiFusion~\cite{artifusion}  &   $30.71$  & $250$ \\ 
\hline
\textbf{Ours}   &  $\bm{0.8341}$ & $\bm{50}$ \\ 
\bottomrule
\end{tabular}
\end{table}

\subsection{Implementation details}
We implement the proposed LatentArtiusion in Python 3.10 and PyTorch 2.1 with Diffusers~\cite{diffusers}.
Hyperparameters are set as follows: batch size is 16, learning rate is $1\times 10^{-4}$ with the Adam optimizer, the downsample factor of the VAE is set to 8, linear variance scheduler is used, and the number of total timesteps is set to 50.

\subsection{Datasets}
We use a subset of Camelyon17~\cite{CAMELYON} comprised of 2445 artifact-free images with the resolution of $256 \times 256$ as the training set \footnote{Available at \url{https://github.com/zhenqi-he/ArtiFusion}.}. To evaluate the performance and robustness of our model, we test on two different data sources containing multiple cell types. (1) Histology: a public histology dataset \footnote{Available at \url{https://github.com/zhenqi-he/transnuseg}.}, which is widely used for nuclei segmentation task~\cite{clusterseg,transnuseg}, consists of $462$ artifact-free images collected from multiple patients with densely clustered nuclei. (2) MoNuSAC 2020~\cite{monusac2020}: another public dataset of $154$ artifact-free tissue images scanned from various organs. In the two test sets, we manually synthesize artifacts into original artifact-free images to obtain paired artifact and artifact-free images for performance evaluation.

\begin{figure*}[t!]
  \includegraphics[width=0.95\linewidth]
  {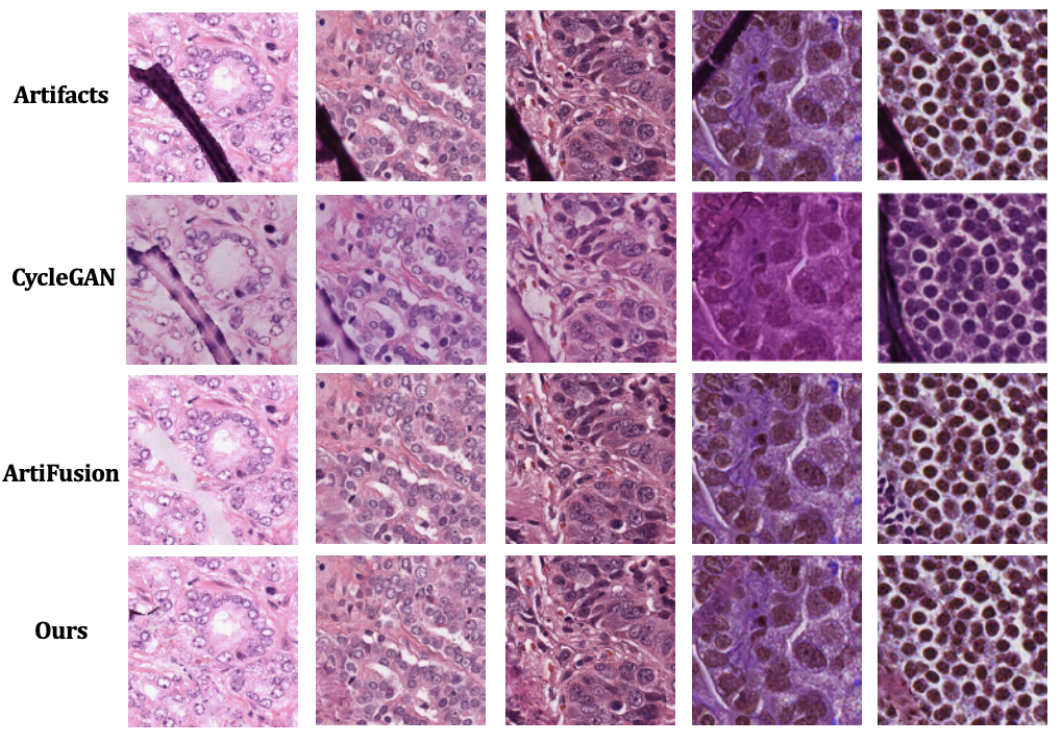}
  \centering
  \caption{
  Qualitative comparison of the restoration by CycleGAN~\cite{CycleGAN2017}, pixel-level ArtiFusion~\cite{artifusion}, and \texttt{Ours}. The first three images are selected from Histology Dataset~\cite{transnuseg} and the last two are from MoNuSAC2020~\cite{monusac2020}.}
  \label{fig:results}
\end{figure*}

\subsection{Comparison and Results}
To quantitatively analyze the restoration performance and computational efficiency, we measure the similarities between unprocessed clean images and restored artifact images using the following evaluation metrics, $L_2$ distance (L2) over the whole image, the mean-squared error (MSE) specifically focused on the artifact region, structural similarity index (SSIM)~\cite{SSIM}, peak signal-to-noise ratio (PSNR)~\cite{PSNR} and feature-based similarity index (FSIM)~\cite{fsim}. Additionally, we measure the time costs during inference. We compare our model with the state-of-the-art models, CycleGAN~\cite{CycleGAN2017} and ArtiFusion~\cite{artifusion}. The quntitative comparisons are shown in Table~\ref{tab:result} and Table~\ref{tab:cost}. Visualized results are demonstrated in Fig.~\ref{fig:results}.

Quantitative comparisons across two distinct data sources highlight the success of our proposed framework in effectively reconstructing tissue structures within artifact regions while maintaining high computational efficiency.
The substantial performance gaps observed across all evaluation metrics between CycleGAN~\cite{CycleGAN2017} and our method corroborate that our latent regional denoising algorithm excels in selectively reconstructing the artifact region while successfully preserving both color and structural information in non-artifact regions.
In contrast to the previous pixel-level approach of ArtiFusion~\cite{artifusion}, denoising in the latent space accelerates the inference time by more than $30$ times, resulting in a substantial reduction from $30$ seconds to just $0.8$ seconds. Moreover, fewer reconstruction steps are required, leading to significant savings in computational resources. 

Overall, our method outperforms the other methods in all metrics across different datasets of various nuclei types collected from multiple organs. At the same time, our method obtains a significant increase by $30\times$ and $1.2 \times$ respectively in the inference time compared with ArtiFusion~\cite{artifusion} and CyclyGAN~\cite{CycleGAN2017}. Moreover, our proposed LatentArtiFusion achieves $80\%$ computational reduction in the reconstruction stage.

\begin{table*}[t]
\centering
\caption{
Quantitative comparison in the downstream tissue classification task. `Clean' indicates unprocessed original images, `Artifacts' means synthetic artifact images, and `Restored w [MODEL]' indicates that the images are restored by the corresponding model.
We evaulate the classification accuracy (\%) on the test set with different network architectures including ResNet~\cite{ResNet}, ConvNeXt~\cite{ConvNeXt}, CSPNet~\cite{wang2020cspnet},
 RexNet~\cite{rexnet} and EfficientNet~\cite{tan2020efficientnet}.
}
\label{tab:classification}
\resizebox{\linewidth}{!}{ 
\begin{tabular}{l|c|c|c|c|c}
\toprule
Settings &  ResNet18   & ConvNeXt\_small  & CSPDarkNet53 & RexNet100  & EfficientNetB0   \\ 
\hline
Clean  & $95.529$ & $95.738$ &    $94.382$  & $95.487$   &  $95.808$   \\
\hline
\hline
Artifacts &$80.302$ & $86.533$ &  $82.210$     &  $90.446$  & $90.626$  \\
\hline
Restored w CycleGAN  & $86.326$ & $87.201$& $86.429$ & $90.776$  & $91.811$\\
\hline
Restored w ArtiFusion  & $92.376$ &  $91.416 $&  $89.792$   &  $92.310$    &  $94.232$ \\
\hline
Restored w \textbf{Ours}  & $\bm{93.201}$ &  $\bm{93.042} $&  $\bm{91.117}$   &  $\bm{93.001}$    &  $\bm{94.694}$ \\
\hline
\end{tabular}
}
\end{table*}
\subsection{Downstream Classification Evaluation}
To validate the effectiveness of our framework in reducing the misdiagnosis rate, we further conduct a downstream classification task. We utilize the publicly available NCT-CRC-HE-100K tissue classification dataset for training and the CRC-VAL-HE-7K dataset for testing. The training set comprises a total of $100,000$ samples, while the test set consists of $7,180$ samples. Various classification models are employed to ensure the robustness of the evaluation across different model architectures. After training all classification models on the training set, we assess their performance on the test data under three different conditions: (a)  clean images, (b) images with synthetic artifacts, and (c) artifact images restored by a specific method. Note that the classification result on clean images serves as the upper bound, while the test result on artifact images represents the lower bound. The quantitative results are shown in Table~\ref{tab:classification}.

A substantial decrease of up to $15\%$ in the classification results across all models is observed when comparing unprocessed data to data with artifacts, demonstrating the disruptive impact of artifacts on deep-learning approaches. The considerable performance gap between CycleGAN~\cite{CycleGAN2017} and diffusion-based methods further validates the superiority of the denoising diffusion model in histological artifact restoration.
Crucially, artifact images restored by our method exhibit the highest classification performance across various model architectures when compared with artifact images reconstructed by CycleGAN~\cite{CycleGAN2017} and ArtiFusion~\cite{artifusion}. Through restoration by our framework, the classification accuracy is improved from approximately $80\%$ to a remarkable $93\%$, which is very close to the classification result using clean images. This underscores the practical effectiveness of our framework in downstream tasks.

\section{Conclusion}
In this paper, we have presented an efficient and effective framework, LatentArtiFusion, for histological artifact restoration, which subtly integrates VAE and DDPM, resulting in a strong model for artifact restoration in a lower-dimensional latent space, leading to a significant increase in computational efficiency. 
Further, we have introduced a regional denoising algorithm in the latent space that successfully preserves the tissue and nuclei information of non-artifact regions. 
The experimental results demonstrate the superiority of our method over previous works across multiple datasets.

\subsubsection{\discintname}
The authors have no competing interests to declare that are
relevant to the content of this article. 

\bibliographystyle{splncs04}
\bibliography{Paper-0041}

\end{document}